\documentclass[onecolumn,preprintnumbers,amsmath,prc,amssymb]{revtex4}

\usepackage{graphicx}
\usepackage{dcolumn}
\usepackage{bm}
\usepackage{amsmath}

\begin{document}

\title{Bose-Einstein Condensation  of  Quantum Hard-Spheres as a Deposition Phase Transition  and New Relations Between 
Bosonic and Fermionic Pressures}

\author{Kyrill A. Bugaev$^{1,2}$, Oleksii I. Ivanytskyi$^{1,3}$, Boris E. Grinyuk$^{1}$  and Ivan  P. Yakimenko$^{4}$}

\affiliation{ $^1$Bogolyubov Institute for Theoretical Physics,
Metrologichna str. 14-B, Kyiv 03680, Ukraine}
\affiliation{$^2$Department of Physics, Taras Shevchenko National University of Kyiv, 03022 Kyiv, Ukraine
}
\affiliation{$^{3}$
CFisUC, Department of Physics, University of Coimbra, 3004-516 Coimbra, Portugal}
\affiliation{$^{4}$
Department of Physics, Chemistry and Biology (IFM), Link\"oping University, SE-58183 Link\"oping, Sweden}

\begin{abstract}
We investigate the phase transition of Bose-Einstein particles with the hard-core repulsion in the grand canonical ensemble within the Van der Waals approximation. 
It is shown that the pressure of  non-relativistic Bose-Einstein  particles is mathematically equivalent  to the  pressure  of simplified version of the statistical multifragmentation model 
of nuclei with  the  vanishing surface tension coefficient and the Fisher exponent $\tau_F = \frac{5}{2}$, which  for such parameters has the 1-st order phase transition. The found similarity of these equations of state  allows us to show that within the present approach  the high density phase of Bose-Einstein  particles is a classical macro-cluster  with vanishing entropy at any temperature which, similarly to the classical hard spheres,  is a kind of solid state. To show this we establish new relations which allow us  to identically   represent the pressure of Fermi-Dirac   particles  in terms of  pressures of  Bose-Einstein  particles of two sorts.
\\

\end{abstract}

\maketitle

\section{Introduction}

The phenomenon  of Bose-Einstein (BE) condensation is, probably,  one of the most striking manifestation of  collective quantum effects \cite{Isihara,Huang}.  
Due to its great importance  the phase transition (PT) of BE condensation in the ideal gas is discussed in all textbooks on statistical mechanics. 
In the wast majority of these textbooks it is written that the BE condensation of ideal gas  is the 3-rd  order phase transition (see, for instance, \cite{Isihara}), although
in the  famous book   \cite{Huang} (see the section 12.3 for details) it is argued that the BE condensation is the 1-st order PT
 between liquid and gas.  The main question we answer here is what kind of PT is the BE condensation in the quantum system with  the simplest interaction, namely  with the hard-core repulsion? 
 In all textbooks it is written that the BE condensate is the group of particles with zero momentum. However, the question is what is it? Is it  a  liquid or a solid?

 In what follows we demonstrate that the pressure of  the  
 non-relativistic BE   particles with the hard-core interaction taken  in the  Van der Waals (VdW) approximation can be identically reduced to the one of  the simplified version of  statistical multifragmentation model (sSMM) \cite{Bondorf} with a vanishing surface tension of the constituents (see below).  
 This exactly solvable model was formulated in \cite{SMM0} and  solved  exactly  in \cite{KABsmm1, KABsmm2,LFT1,Reuter08}, while its new and more realistic generalization can be found in \cite{KABsmm3}. 
 Although  the  sSMM  \cite{SMM0, LFT1, KABsmm1, KABsmm2, Reuter08, KABsmm3} lacks   the Coulomb interaction between the nuclei and the asymmetry energy of nuclei, its exact analytical solution established both in the 
thermodynamic limit \cite{KABsmm1, KABsmm2, KABsmm3} and for finite volumes \cite{LFT1, Reuter08} is able to qualitatively describe  the main properties of the nuclear  liquid-gas PT.  

The mathematical similarity between  the VdW EoS of BE hard spheres and the sSMM allows us to show that the high density phase of  BE  particles with hard-core repulsion is a classical macro-cluster  
which, similarly, to the classical hard-spheres is 
a solid state  \cite{Simple_Liquids,Mulero} and not a liquid as it was argued in K. Huang  book   \cite{Huang}.  In our analysis we also
analyzed the pressure of  Fermi-Dirac (FD) particles  with the hard-core repulsion which in many respects is similar to the one of sSMM, although  it does not have the 1-st order PT.
This analysis allows us to find out  some new  relations between the 
pressures of  BE  and FD particles with the hard-core repulsion, which help us to
demonstrate that the macro-cluster of BE particles  is, indeed, a classical object. 
The found relations  allow us to  clearly demonstrate   under what   conditions 
the FD particles with the hard-core repulsion can have the first order phase transition.

The work is organized as follows. In Sect. II we analyze the pressure of BE and FD particles with the hard-core repulsion in the VdW approximation  in a form convenient  for the grand canonical ensemble. Sect. III is devoted to  discussion of the properties of the macro-cluster with the help
of the BE-FD decomposition identities which identically   represent the pressure of FD particles   in terms of two  BE pressures. Our conclusions are given in Sect. IV.

\section{BE condensation as  the 1-st order Phase Transition}

The equation of state (EoS) of hard-spheres with BE or FD statistics in the grand canonical ensemble variables under the Van der Waals approximation for the hard-core repulsion can
be obtained either analyzing  the free energy of  the Van der Waals gas in canonical ensemble \cite{Qvdw1,Qvdw2}
or more rigorously from the quantum partition function in the grand canonical ensemble \cite{GenISCT2019}.
In the grand canonical variables it has the form 
\begin{eqnarray}\label{Eq1}
&&\hspace*{-3.5mm}p_\pm = p^{id}_\pm (T, \nu) \equiv \pm  T g \hspace*{-1.5mm} \int \hspace*{-1.5mm}  \frac{d^3 k }{(2 \pi \hbar)^3}   \ln  \left[1 \pm  \exp\left[ \frac{\nu - e(k)}{T} \right] \right] , ~ \rm where\quad \nu \equiv  \mu - 4 V_0\,  p_\pm  ,\quad
\end{eqnarray}
where the lower sign  is for  the BE statistics, while the upper sign is for the FD one.  
Here $T$ is  temperature of the system, $\mu$ is  its chemical potential, $\nu$ is an effective chemical potential,   $g$ is the number of spin-isospin states (degeneracy factor), $m$ is the mass of particle,
$V_0 = \frac{4}{3} \pi R^3$ is the ``eigen volume`` of particle, and $R$ is the half of the minimal interaction range of the hard-core   potential 
$U(r)$
of a one component system (with a single hard-core radius)
\begin{eqnarray}\label{Eq2}
U(r) = \left\{ \begin{array}{lc}
 0, & |r| > 2 R  ,\\
  \infty, & |r| \le 2 R .  \\
\end{array} \right. 
\end{eqnarray}
The potential $U(r)$ acts in a simplest possible way: (i)
if two particles 1 and 2, for definiteness,  do not interact, i.e.   the distance between them  $|r| > 2 R$ is   larger, than two hard-core radii $R$, then $U(r)=0$ and, therefore,   their total energy is the sum of their  single-particle (kinetic) energies $e_1$ and $e_2$; (ii) if  these  two particles interact, then $|r| = 2 R$ and $U(|r| =2R) = \infty$, but such configurations 
do not contribute to partition (and all thermodynamic functions), since they 
are suppressed by
the statistical operator $\exp\left[- \frac{\hat H_{hc}}{T} \right]$ due to an infinite potential  energy (here $\hat H_{hc}$ denotes  the Hamiltonian of the system). As a result, the total energy of the particles with the hard-core repulsion equals to the sum of  their  single-particle  (kinetic) energies  and this allows one to find the pressure (\ref{Eq1}) directly from the quantum partition function.  In other words, the particles with the hard-core interaction behave as an ideal quantum gas. 

This is a important property of this EoS which leads to a well-known  practical consequence, namely that  the energy per particle  coincides with the one of the ideal gas. Due to this property the sophisticated equations of state with the hard-core repulsion, known as the hadron resonance gas model,  are very successfully used to   describe the multiplicities of hadrons \cite{IST2018,IST2018b,KABugaev19} and light (anti-, hyper)nuclei \cite{IST2020a,IST2020b} which are measured in the high energy nuclear collisions  and to get a reliable thermodynamic information  about next to the last stage of such collisions.

For further analysis it is convenient to introduce the auxiliary functions
\begin{eqnarray}\label{Eq3}
&&\hspace*{-7.7mm} {\cal F}_{\pm} (p) \equiv  T  \sum\limits_{l=1}^{K_{max}} \frac{(\mp1)^{(l+1)}}{l} n^{id}_0 \left[\frac{T}{l}, \nu (p)  \right], ~~ \Rightarrow \quad \rm Eq. \, (1) ~becomes \quad p_\pm =  {\cal F}_\pm (p_\pm) \, , \\
&&\hspace*{-7.7mm} n^{id}_0 \left[\frac{T}{l}, \nu   \right]  = \int  \hspace*{-1.2mm}\frac{g\, d^3 k   }{(2 \pi \hbar)^3}  e^{\frac{l \left[ \nu -\sqrt{m^2+\vec k^2} \right] }{T} } \simeq 
\label{Eq4}
  \int  \hspace*{-1.2mm}\frac{g\, d^3 k   }{(2 \pi \hbar)^3}  e^{\frac{l \left[ \nu - m - \frac{k^2}{2 m } \right] }{T} } = g \left[ \frac{m \, T}{ 2 \pi\, l  \,\hbar^2} \right]^\frac{3}{2} e^{\frac{l \left[ \nu - m\right] }{T} } ,~
\end{eqnarray}
where  the particle number density of Boltzmann point-like particles with temperature $T$ and  chemical potential  $\nu $ is denoted as $n^{id}_0 \left[T, \nu   \right] $ and the upper limit of sum in Eq. (\ref{Eq3}) is $K_{max} \rightarrow \infty$.  To avoid the unnecessary complexity  in our derivations through out this work   we  regard the limit  $K_{max}= 2K +1 \rightarrow \infty$ strictly  in this sense.  For the BE statistics  (sign $-$ in Eqs. (\ref{Eq3}) and (\ref{Eq4})) it is not important, but  it is very important for  the case of FD statistics (sign $+$ in Eqs. (\ref{Eq3}) and (\ref{Eq4})) . 

The function ${\cal F}_{\pm}$ in (\ref{Eq3}) is, apparently,  obtained   by expanding the $\ln$-function in Eq. (\ref{Eq1}). 
For large values of $l \gg 1$ the inequality $l m \gg T$ is valid for any non-vanishing mass  $m$ and, therefore, in this case one can use the non-relativistic  approximation  in the left hand side momentum integral in  Eq. (\ref{Eq4}) and get the right hand side expression (\ref{Eq4}). However, for convenience  we will use 
such an approximation for any $l \ge 1$, assuming that considered temperatures are very low compared to the particle mass,
i.e. $m \gg T$.   Moreover, in what follows we will always use the non-relativistic approximation for particle energy, unless it is specified explicitly. 

To  make a direct comparison with the sSMM  \cite{SMM0,KABsmm1, KABsmm2, LFT1, Reuter08}  
we  explicitly write Eq. (\ref{Eq1}) for the BE statistics $(a=-1)$
\begin{eqnarray}\label{Eq5}
%
\hspace*{-5.5mm} p_- &=&   T \, g \left[ \frac{m \, T}{ 2 \pi  \hbar^2} \right]^\frac{3}{2}  
\sum\limits_{k=1}^{K_{max}} \frac{ (-a)^{(k+1)}  }{  k^\frac{5}{2} } 
 \exp  \left[ \frac{k  (\mu - m - 4 V_0\, p_- )   }{T}  \right]  , ~
\end{eqnarray}
using Eq. (\ref{Eq3}) and the right hand side  Eq. (\ref{Eq4}).
Comparing Eq. (\ref{Eq5})  with Eq. (15) from Ref.  \cite{KABsmm1}, one can see  that  the  pressure  of BE hard spheres 
 is  mathematically absolutely  equivalent to the sSMM with the ``volume`` $4 k V_0$ of $k$-nucleon  nuclei,  
with   the vanishing surface tension of all nuclei  and  with the Fisher exponent $\tau_F = \frac{5}{2}$ (or for the index $\tau \equiv \tau_F + \frac{3}{2} = 4$ in terms of Refs. \cite{KABsmm1,KABsmm2}). 

Due to the mathematical similarly to the sSMM, using the exact solution of sSMM \cite{KABsmm1, KABsmm2, Reuter08}  one can immediately conclude that  Eq. (\ref{Eq5})  describes two phases: 
the gaseous phase   $p_g =  p_- (T, )$ for the low densities defined by the inequality  $\mu < \mu_c(T) $, and high density 
phase pressure $p_s =  \frac{(\mu-m)}{4 V_0 }$  for  $\mu > \mu_c(T) $.  
According to the Gibbs criterion the  PT  occurs, if  the pressures of two phases are equal, i.e.  
$p_g (T, \mu_c) = p_s (T, \mu_c)$. This equation   defines the phase equilibrium curve $\mu = \mu_c(T) $ of the 1-st order PT. 

At the PT curve $\mu = \mu_c(T) $ the effective chemical potential becomes
\begin{eqnarray}\label{Eq6}
\hspace*{-5.5mm} \nu_c &=& \mu_c - 4 V_0 \, p_g (T, \mu_c) = \mu_c - 4 V_0 \, p_s (T, \mu_c) \equiv m \,.~
\end{eqnarray}
Using this result  one can identically rewrite the pressure at PT curve as 
\begin{eqnarray}\label{Eq7}
\hspace*{-5.5mm} p_{c-} &=&   T  \,  g \left[ \frac{m \, T}{ 2 \pi  \hbar^2} \right]^\frac{3}{2}  \sum\limits_{k=1}^{K_{max}} \frac{ 1  }{  k^\frac{5}{2} } \underbrace{=}_{K_{max} \rightarrow \infty}
 \frac{ T  \,  g}{\Gamma\left[ \frac{5}{2}\right]}
 \left[ \frac{m \, T}{ 2 \pi  \hbar^2} \right]^\frac{3}{2} \int\limits_0^\infty \frac{t^\frac{3}{2}}{e^t-1} dt 
  , ~ 
\end{eqnarray}
where we used the integral representation of the Riemann $ \zeta\left[ \frac{5}{2}\right]$-function \cite{Prudnikov}. Here $\Gamma  (n+1) = n!$ is the usual gamma-function.  Taking $t = \frac{\omega}{T}$ in the integral in Eq. (\ref{Eq7}), one recovers the traditional representation of pressure as an integral over the particle energy $\omega$ \cite{Isihara}.

Although the critical pressure (\ref{Eq7}) coincides with the one obtained usually for the point-like particles   \cite{Isihara},
the particle number density of gas  $n_-$ is modified due to the presence of hard-core interaction. Using the particle number density 
of   the gas of  point-like particles $n^{id}_{-} (T, \nu)$ one can write
\begin{eqnarray}\label{Eq8}
\hspace*{-5.5mm} n^{id}_{-} (T, \nu) &\equiv & \frac{\partial p^{id}_{-} (T, \nu)}{\partial \nu} =     g \left[ \frac{m \, T}{ 2 \pi  \hbar^2} \right]^\frac{3}{2}  \sum\limits_{k=1}^{K_{max}} \frac{ 1  }{  k^\frac{3}{2} } 
 \exp  \left[ \frac{k  (\nu-m)   }{T}  \right]    , ~ \\
 \label{Eq9}
\hspace*{-5.5mm} n_{-} (T, \nu) &\equiv & \frac{\partial p^{id}_{-} (T, \nu)}{\partial \mu} = \frac{n^{id}_{-} (T, \nu)}{1+ 4V_0 n^{id}_{-} (T, \nu)}  .
\end{eqnarray}
From  Eq.  (\ref{Eq9}) one  can see that at the PT curve the particle number density of the gas  is smaller than the
particle number density of the dense phase, since
\begin{eqnarray}\label{Eq10}
\hspace*{-5.5mm} n_{-} (T, \nu_c) & = &  \frac{n^{id}_{-} (T, \nu_c)}{1+ 4V_0 n^{id}_{-} (T, \nu_c)}  < n_s \equiv \frac{\partial p_s}{\partial \mu} = \frac{1}{4 V_0}~ ,
\end{eqnarray}
and, hence,  for any finite temperature $T$ the particle number density of point-like particles $n^{id}_{-} (T, \nu_c)$ is finite too. Therefore, the particle number density  of gaseous phase is 
smaller than the one of the high density phase as indicated by the inequality (\ref{Eq10}).
As a result, the BE PT is of the 1-st order. 

Substituting into Eqs. (\ref{Eq8}) and (\ref{Eq9}) the value  $\nu=\nu_c$ one can get the temperature of BE condensation as
\begin{eqnarray}\label{Eq11}
\hspace*{-5.5mm} T^{BE}_c & =&\frac{2 \pi \hbar^2}{m} \left[\frac{1}{g \zeta\left[ \frac{3}{2}\right]}\cdot  \frac{n_-}{1- 4 V_0 n_-} \right]^\frac{2}{3}  . ~ 
\end{eqnarray}
Note that for large values of the excluded volume $V_0$ and high particle number densities $n\rightarrow \frac{1}{4 V_0}$
the hard-core repulsion  may essentially increase
the value of the PT  temperature and make it more realistic compared to  the traditional  estimate
obtained for the point-like particles \cite{Isihara}, i.e. if one takes  the limit  $V_0 \rightarrow 0$ on the  right hand side of  Eq. (\ref{Eq10}). 

It is necessary to stress that the above results are generic in a sense that one can 
consider the effective values of  degeneracy factor  $g \rightarrow g^{eff}$ and the one of 
excluded volume $4 V_0 \rightarrow V_0^{eff}$ which correspond to a more realistic EoS  than the VdW EoS  and which is able to reproduce the pressure  of quantum particles beyond the second virial coefficient approximation at least in   some (even in a  narrow)  
range of thermodynamic parameters.

Since we are also  interested in analyzing  the  case  of FD particles, we would like to 
 obtain the above  result  using a different approach, namely without referring to the sSMM results of Refs. \cite{KABsmm1,KABsmm2,LFT1}.
First we consider  the  limit $\mu =  \rightarrow \infty$ in Eq. (\ref{Eq5}) for very large, but finite values of $K_{max}$.  Apparently, this limit should correspond to the dense phase  of our  EoS. Then  in this limit $V_0 p_s /T \gg 1 $ and for $\mu  >  m + 4 V_0  p_s$ the leading terms of Eq. (\ref{Eq5}) for $a=-1$ can be written as
\begin{eqnarray}\label{Eq12}
&&\hspace*{-15.5mm}  \ln \left[\frac{p_s K^\frac{5}{2}_{max}}{T \phi(T) }\right]    \simeq 
 K_{max} \left[ \mu - ( m + 4 V_0 p_s)\right] . ~
\end{eqnarray}
Here   the thermal density of the gas of classical hard spheres is denoted as $\phi(T) = g \left[ \frac{m \, T}{ 2 \pi  \hbar^2} \right]^\frac{3}{2}$. In deriving Eq. (\ref{Eq5})  we have chosen 
the large values of  chemical potential $\mu > \mu_c$, which are not allowed in the thermodynamic limit, but for finite systems they can be used \cite{LFT1,Reuter08}.
Now from Eq. (\ref{Eq12})  one can see that for $K_{max} \rightarrow \infty$ the logarithmic correction disappears and the pressure of dense phase  $p_s = \frac{\mu -m}{4 V_0} $  acquires a familiar  form.

In order to show that the EoS (\ref{Eq5}) for $a=-1$ has the 1-st order PT we  examine the derivative $D^1p_- \equiv T \frac{\partial p_-}{\partial \rho^{id}_-}$.  Hereafter to avoid a confusion  we will  distinguish the particle number density of point-like particles  as the function given
by the right hand side of  Eq. (\ref{Eq8}) and the same quantity as the independent variable $\rho^{id}_-$. The derivative
$D^1p_- $ is more convenient to employ for the spinodal instability point of the gas than the derivative $\frac{\partial p_-}{\partial n_-}$, since its expression is simpler.  Note that vanishing of the spinodal instability point of the gas taken at  the given isotherm signals about the 1-st order PT \cite{Isihara}. Indeed,  the expression for $D^1p_- $  
\begin{eqnarray}\label{Eq13}
& D^1p_- \equiv  T \frac{\partial p_-}{\partial \nu}   \frac{\partial \nu}{\partial \rho^{id}_-} = \left[ \sum\limits_{k=1}^{K_{max}} \frac{ 1  }{  k^\frac{3}{2} } 
 \exp  \left[ \frac{k  (\nu - m )   }{T}  \right]  \right]   \left[ \sum\limits_{k=1}^{K_{max}} \frac{ 1  }{  k^\frac{1}{2} } 
 \exp  \left[ \frac{k  (\nu - m )   }{T}  \right]  \right]^{-1} , ~~
\end{eqnarray}
shows that, if  the effective chemical potential  $\nu =\mu - 4 V_0\, p_- $ approaches the value $\nu=m$, then  the derivative $\frac{\partial \rho^{id}_-}{\partial \nu}  \equiv \frac{\partial n^{id}_-}{\partial \nu} \rightarrow \infty$ diverges for $K_{max} \rightarrow \infty$ and, hence,  in this limit $ D^1p_- =0$. Thus, we have found that
the spinodal instability point of the gas  of BE hard spheres coincides with the PT curve.  

Now we turn to the analysis of the FD particles with the hard-core repulsion.  For $\nu \le m $ the pressure of such particles $p_+ $ and its $\nu$-derivative  can be explicitly written as
\begin{eqnarray}\label{Eq15}
%
\hspace*{-5.5mm} p_+ &=&   T \, g \left[ \frac{m \, T}{ 2 \pi  \hbar^2} \right]^\frac{3}{2}  
\sum\limits_{k=1}^{K_{max}} \frac{ (-1)^{(k+1)}  }{  k^\frac{5}{2} } 
 \exp  \left[ \frac{k  (\nu - m )   }{T}  \right]  \Bigg|_{\nu=\mu  - 4 V_0\, p_+} , ~\\
\label{Eq16}
\hspace*{-5.5mm} n^{id}_+ &\equiv &  \frac{\partial p_+}{\partial \nu} =   g \left[ \frac{m \, T}{ 2 \pi  \hbar^2} \right]^\frac{3}{2}  
\sum\limits_{k=1}^{K_{max}} \frac{ (-1)^{(k+1)}  }{  k^\frac{3}{2} } 
 \exp  \left[ \frac{k  (\nu - m)   }{T}  \right] \Bigg|_{\nu=\mu  - 4 V_0\, p_+}  .
\end{eqnarray}
A similarity with the sSMM can be more clearly seen for $\nu =\mu  - 4 V_0\, p_+ \rightarrow m$, if in the sum (\ref{Eq15})  one adds the even terms to the preceding odd ones 
\begin{eqnarray}\label{Eq16n}
%
\hspace*{-5.5mm} p_+ &=&   T \, g \left[ \frac{m \, T}{ 2 \pi  \hbar^2} \right]^\frac{3}{2}  
\left[
\sum\limits_{k \in odd }^{K_{max}-2} \frac{  \exp  \left[ \frac{k  (\nu - m )   }{T}  \right] }{  k^\frac{5}{2} } 
  \left[ 1- \frac{k^\frac{5}{2}}{(k+1)k^\frac{5}{2}}  \exp  \left[ \frac{  (\nu - m )   }{T}  \right]\right] +  \frac{  \exp  \left[ \frac{K_{max}  (\nu - m )   }{T}  \right] }{  K_{max}^\frac{5}{2} } 
 \right] \simeq   ~\\
 \hspace*{-5.5mm}  & \simeq & T \, g \left[ \frac{m \, T}{ 2 \pi  \hbar^2} \right]^\frac{3}{2}  
\left[ \sum\limits_{k \in odd }^{K_{max}-2} \frac{ 5 }{2  \, k^\frac{7}{2} } 
 \exp  \left[ \frac{k  (\nu - m )   }{T} - \frac{5}{2 k} \right] +  \frac{  \exp  \left[ \frac{K_{max}  (\nu - m )   }{T}  \right] }{  K_{max}^\frac{5}{2} } \right], ~
 \label{Eq17n}
\end{eqnarray}
where we expanded the binomial $(k+1)^\frac{5}{2}$ keeping two leading terms and approximated the ratio $k^\frac{5}{2}/(k+1)^\frac{5}{2} \simeq \exp[-\frac{5}{2k}]$. Evidently, this approximation is suited for $k \gg 1$, but for qualitative analysis it is very convenient, since in the vicinity of PT the main role is played by the largest cluster.  Eq. (\ref{Eq17n}) shows that, apart from the term with $k= K_{max}$,   in the left vicinity of the point 
$\nu \rightarrow m - 0$ the EoS for FD particles with the hard-core repulsion is similar to the sSMM
for the clusters of the odd number of constituents which have
 the Fisher exponent $\tau_F = \frac{7}{2}$ and a vanishing value of  surface tension coefficient.

Apparently, from Eqs.  (\ref{Eq15})  and (\ref{Eq17n})  one can also derive Eq.  (\ref{Eq12}) and 
 establish the pressure of dense phase  $p_s = \frac{\mu-m}{4 V_0}$ similarly to the case of BE particles. However,  the 
 derivative 
\begin{eqnarray}\label{Eq17}
\frac{\partial \rho^{id}_+}{\partial \nu}  \equiv   \frac{\partial^2 p_+}{\partial \nu^2} = \frac{g}{T} \left[ \frac{m \, T}{ 2 \pi  \hbar^2} \right]^\frac{3}{2}  
\sum\limits_{k=1}^{K_{max}} \frac{ (-1)^{(k+1)}  }{  k^\frac{1}{2} } 
 \exp  \left[ \frac{k  (\nu - m)   }{T}  \right] ,
\end{eqnarray}
with respect to the effective chemical potential $\nu = \mu - 4 V_0\, p_+$ 
is finite  for $\nu=m$, since, in contrast to the case of BE particles,   the sum staying in Eq. (\ref{Eq17}) converges in the limit $K_{max}\rightarrow \infty$.
Indeed, with the help of integral representation of the Riemann $\zeta$-function \cite{Prudnikov} for $\nu=m$ one finds 
\begin{eqnarray}\label{Eq18}
\hspace*{-5.5mm}  \sum\limits_{k=1}^{\infty} \frac{ (-1)^{(k+1)}  }{  k^\frac{1}{2} } =  \frac{1}{\Gamma\left[ \frac{1}{2}\right]} \int\limits_0^\infty \frac{t^{-\frac{1}{2}} }{e^t+1} dt \simeq 0.6049
 , ~ 
\end{eqnarray}
and,  therefore,  the derivative $D^1p_+ \equiv T \frac{\partial p_+}{\partial \rho^{id}_+}$ does not vanish for $\nu=m$ and, hence, there is no 1-st order PT in this case. 

In our opinion this is a very simple and  good example that the presence  of a  macro-cluster  with 
the finite probability  in a finite system is a necessary, but not a sufficient condition of the 1-st order PT existence in such a system. We believe this is an important message to be taken into account by the authors of Refs. \cite{Francesca2014,DasGupta2018} who consider  the presence and gradual disappearance of the macro-cluster as a signal of the 1-st order nuclear liquid-gas PT
in finite systems. The whole point is that in finite systems the macro-cluster of maximal size can appear as
the metastable state of finite probability not only for the 1-st order PT, but also for the 2-nd order PT or even for the cross-over \cite{LFT1,Reuter08}. 
The present analysis  once more shows one 
that for vanishing surface tension coefficient  the value of the Fisher exponent $\tau_F$ defines the PT order \cite{Reuter08}.

One can readily check that all the results on PT existence remain valid, if one uses the relativistic expression for particle energy, i.e. if one makes a replacement $m + \frac{k^2}{2 m} \rightarrow \sqrt{m^2 + k^2}$. However, in this case the BE condensation does not look mathematically identical
to the sSMM and, hence, the corresponding analysis is not made here. 

\section{Decomposition identity between bosonic  and fermionic pressures}

Apart from the formal difference between the EoS of the BE and FD particles we would like to 
understand (i)
whether our interpretation of the appearance of classical macro-clusters is correct, and (ii) under what circumstances the appearance of macro-cluster can be associated with the 1-st order PT  in the system of FD particles. 
Indeed, an absence of the 1-st order PT  in the EoS of FD particles with the hard-core repulsion may 
question the validity of  our hypothesis about the classical macro-clusters  existence and, therefore,  one may think that BE condensation leads to an appearance of quantum macro-cluster with BE statistics, while 
the quantum macro-cluster with FD statistics cannot be formed due to some reason, namely due to the Pauli blocking principle. 

To demonstrate the validity of our hypothesis we consider a peculiar mathematical identity between the BE and FD  pressures which we call {\it  a BE-FD decomposition identity}
%
\begin{eqnarray}\label{EqB11}
&&\underbrace{- T g \hspace*{-1.5mm} \int \hspace*{-1.5mm}  \frac{d^3 k }{(2 \pi \hbar)^3}   \ln  \left[{\textstyle 1 - \exp\left[ \frac{\nu - \sqrt{m^2 +k^2}}{T}  \right] }\right] }_{p_B\left(\frac{\nu}{T}, m, g\right)}\equiv 
\nonumber  \\
%
&& \equiv T g \hspace*{-1.5mm} \int \hspace*{-1.5mm}  \frac{d^3 k }{(2 \pi \hbar)^3}  \left\{
 \underbrace{ 
\ln  \left[{\textstyle 1 + \exp\left[ \frac{\nu - \sqrt{m^2 +k^2}}{T}  \right] } \right]}_{p_F\left(\frac{\nu}{T}, m, g\right)} 
 \underbrace{ -
\frac{1}{8} \ln  \left[{\textstyle 1 - \exp\left[ \frac{2 \nu - \sqrt{4m^2 +k^2}}{T}  \right] } \right] }_{p_B\left(\frac{2\nu}{T}, 2m, 2^{-3}g\right)} \right\} ,~
\end{eqnarray}
which will
 help  us  to understand the appearance of a classical macro-cluster for BE and  FD statistics. The fact that now we do not use the non-relativistic approximation to the particle energy is not important. 
 
 The BE-FD decomposition  identity (\ref{EqB11}) can be obtained in the following sequence of steps: first we note that 
\begin{eqnarray}\label{EqB12}
%
%
%
&& \hspace*{-3.5mm} {\textstyle \ln  \left[ 1 - \exp\left[ \frac{2 \nu - \sqrt{4m^2 + 4k^2}}{T}  \right]  \right]  \equiv 
\ln  \left[ 1 +    \exp\left[  \frac{\nu - \sqrt{m^2 +k^2} }{T}  \right]  \right] +  } 
  {\textstyle  \ln  \left[ 1 - \exp \left[ \frac{ \nu - \sqrt{m^2 +k^2}}{T}  \right]  \right]   }. ~
\end{eqnarray}
Next one can integrate  Eq.  (\ref{EqB12})  over $d^3 k$  with the degeneracy factor $g$ and change a particle momentum  on the left hand side of Eq.  (\ref{EqB12}) as $2 k \rightarrow k$ and in the momentum integral to get a multiplier $\frac{1}{8}$. Finally, interchanging the positions of integrals for  lighter and heavier bosons 
one arrives at  Eq.  (\ref{EqB11}).

Eq.  (\ref{EqB11}) shows one that  for the given values of  $T$ and $\nu$  the pressure of ideal gas of bosons (the upper line of Eq. (\ref{EqB11})) of mass $m$ and degeneracy $g$  can be identically decomposed 
into the sum of two terms. The first pressure corresponds to  the ideal gas of fermions with same mass and degeneracy (the first  term on the right hand side  of Eq. (\ref{EqB11})), while  the second pressure describes the   bosons with the double mass and double 
charge (and  the double excluded volume $V_0$, if $\nu = \mu - 4 V_0 p$ accounts for the effects of hard-core repulsion as above), but with  the reduced  degeneracy $\frac{g}{8}$.  Then the  heavy bosons may be interpreted  as ``pairs`` of fermions. 

Applying the  BE-FD  decomposition identity (\ref{EqB11}) $(n-1)$ times to the pressure $p_B\left(\frac{2\nu_B}{T}, 2m, 2^{-3}g\right)$ of ``pairs``,
 one can identically  extract the contribution of   bosonic macro-cluster ($n \gg 1$) with the mass $2^n m$, the charge $2^n$ and the degeneracy 
 $2^{-3 n}g$ from  the pressure of bosons of mass $m$, charge $1$ and degeneracy $g$ and get 
 the following useful relation
\begin{eqnarray}\label{EqB13}
p_B\left(\frac{\nu_B}{T}, m, g\right) \equiv & p_B\left(\frac{2^n\nu_B}{T}, 2^n m, 2^{-3n}g\right) +  
\sum\limits_{k=0}^{n-1}    p_F \left(\frac{2^k\nu_B}{T}, 2^km, 2^{-3k}g \right) , 
\end{eqnarray} 
where $ p_F \left(\frac{2^k\nu_B}{T}, 2^km, 2^{-3k}g \right)$ denotes the pressure of auxiliary  fermions  with the mass $2^k m$, the charge $2^k$ and degeneracy  $2^{-3 k}g$.  For low temperatures $T\ll m$ one can safely use the non-relativistic approximation 
for the energy of particle. 
Applying  the  identity  (\ref{EqB13})  to the gas pressure of bosons $p_-$ of the EoS considered in the preceding section, i.e. for $\nu_B \le \nu_c$, one can immediately conclude that for $\nu_B < m$  the effective chemical potential of the bosonic  macro-cluster  on the right hand side of  Eq. (\ref{EqB13})  is  $(\nu_B -m) 2^n \rightarrow - \infty$ for $n\gg 1$ and, hence,  such a macro-cluster does not exist for $\nu_B < m$. It is evident,  that the bosonic  macro-cluster  on the right hand side of  Eq. (\ref{EqB13})  does not exist for  $\nu_B = m$ as well, since for $n\gg 1$ its degeneracy 
$2^{-3n}g \rightarrow 0$ vanishes. Apparently, this argument is valid for the case $\nu_B < m$ as well. 
Therefore, in the whole gaseous phase and at the condensation curve of  the  EoS of BE particles
with the hard-core repulsion   considered  in the preceding section  the bosonic macro-cluster
is absent, i.e. for $\nu_B \le m$ one finds 
\begin{eqnarray}\label{EqN23}
p_B\left(\frac{\nu_B}{T}, m, g\right) = & \sum\limits_{k=0}^{\infty}    p_F \left(\frac{2^k\nu_B}{T}, 2^km, 2^{-3k}g \right) ,
\end{eqnarray} 
that the pressure of BE particles can be identically written as an infinite  sum of the pressures of FD particles with certain masses, charges and degeneracies. 
In the preceding section it was shown  that the pressure of  FD particles  with the hard-core repulsion does not have the 1-st order PT and, thus, in the thermodynamic limit  there is no fermionic macro-cluster for each pressure staying  on the right hand side of Eq. (\ref{EqN23}). However, the pressure of BE particles staying on the left hand side of Eq. (\ref{EqN23}) demonstrates the  1-st order PT  of the BE condensation. Therefore, the only 
possible explanation out of this apparent contradiction is that the BE condensation leads to  an appearance of the  classical macro-cluster which is  the sum of individual classical  macro-clusters generated by the set of fermionic pressures that are staying on  the right hand side of Eq. (\ref{EqN23}).

Now it is appropriate  to discuss the properties of the dense phase of  BE  hard spheres within the VdW approximation. 
Since the pressure of dense phase   $p_s = \frac{\mu-m}{4 V_0}$  does not depend on the temperature explicitly, then the  entropy density of dense phase $s_s = \frac{\partial p_s}{\partial T} =0 $ is zero at any temperature, while the  particle number density of this phase  is $n_s = \frac{\partial p_s}{\partial \mu} = \frac{1}{4 \, V_0}$. Furthermore, from the thermodynamic identity 
\begin{eqnarray}\label{Eqn24}
\varepsilon_s = Ts_s + \mu n_s - p_s =  \frac{m}{4 \, V_0}
\end{eqnarray}
one can see that  the energy density $\varepsilon_s$ of the dense phase, indeed, corresponds to the particles at rest which have the highest possible  density within the adopted approximation.  
Therefore, similarly to the  case of classical hard spheres  it is more appropriate  to call this phase as the solid state \cite{Simple_Liquids,Mulero} (since there is not attraction among the particles and the surface tension coefficient is zero).
Furthermore,  it seems  it  is   more appropriate  to consider the BE condensation of particles with the hard-core repulsion as  the deposition PT from a gas to a solid.  
Of course, one has to remember that, on the other hand,  it is a condensate of hard spheres with a vanishing  momentum. 

A mathematical  similarity with the exact solution of sSMM  allows one  to reliably  interpret the BE condensation of hard spheres as the 1-st order PT  in which the gas condenses into a classical macro-cluster  of the size $4V_0 K_{max}$ with $K_{max}\rightarrow \infty$ in the thermodynamic limit.  
Hence, at the PT   curve there should exist the phase boundary.   As it was shown above, formally,  a macro-cluster  corresponds to the term $k= K_{max}$
in the expression for  pressure $p_-$ in  Eq.  (\ref{Eq5}). Therefore,  formally  a macro-cluster  can be considered  as a single classical particle which is at rest.  From Eq. (\ref{Eq5}) one can see that its 
statistical  weight  is  the    Boltzmann one.   Such an interpretation is similar to the sSMM \cite{KABsmm1,KABsmm2} with the difference that in the sSMM
a macro-cluster is a droplet of liquid which has the non-vanishing  surface tension below the critical temperature \cite{Bondorf,KABsmm1,KABsmm2,KABsmm3}  and a finite entropy which vanishes at $T=0$ only. 

 Moreover, considering the EoS (\ref{Eq5}) with the effective values of  $g \rightarrow g^{eff}$ and $4 V_0 \rightarrow V_0^{eff}$
which allow one  at least in the narrow range of thermodynamic parameters to reproduce the realistic EoS of quantum particles  at high densities close $0.45/V_0-0.55/V_0$
and sufficiently high temperature $T$ for which the effects of quantum statistics are not important, one should still have the BE condensation PT on the one hand. On the other hand, 
this should be the region of the deposition phase transitions for the classical hard spheres \cite{Simple_Liquids, Mulero}.  Thus, we again should conclude that at high temperature $T$ the BE condensation of quantum hard spheres should match with the deposition PT of  
the classical hard spheres.

Coming back to the ideal gas of BE particles one should consider the limit $V_0 \rightarrow 0$ in all formulas above. In this limit the high density state has infinite particle number density and, hence, it is inaccessible. However, for any infinitesimally small eigenvolume $V_0$  our conclusions 
about the deposition PT remain valid and, therefore,  the whole argumentation of K. Huang 
in Ref. \cite{Huang} about the BE condensation as the 1-st order PT is correct. Only the K. Huang 
interpretation of this PT as a gas-liquid one   seems to be inconsistent with the modern interpretation 
of the PT of hard spheres. 

At the moment it is not clear, if it is just  a coincidence that at low pressures  the  real gases of  mono- and diatomic molecules, except for  the helium-4 for pressures below 25 atm.,   indeed, demonstrate the deposition PT  under cooling.  Maybe a more realistic EoS of quantum particles can resolve this problem. 

It is remarkable that the BE-FD decomposition identity (\ref{EqB11})  allows one to establish another important interpretation.  The right hand side of  the identity (\ref{EqB11})  corresponds to  pressure of a  mixture of the 
ideal gases of fermions and their pairs (which are the bosons) with the same degeneracy, but with the double mass and double charge, which are taken with the wight $1/8$. The left hand side of  the identity (\ref{EqB11})  shows that such a mixture should experience the 1-st order PT of  BE condensation.   From the famous work of   L. N. Cooper  \cite{Cooper} it is known that the pairing 
of fermions can, indeed, happen under not very restrictive  conditions leading to the BE condensation of fermionic pairs 
and the  BE-FD decomposition identity (\ref{EqB11})  illustrates  such a possibility for a mixture
discussed above. However, for the  appearing of Cooper pairs  the fermions must have an attraction, which is absent in the 
EoS discussed here. 

It is evident that the identity (\ref{EqB12}) is valid for any dimension $D = 1, 2, 3, ...$. Introducing the pressures  of  BE particles (sign $-$) and FD particles (sign $+$) of mass $m$ that have  the chemical potential $\nu$ and temperature $T$
\begin{eqnarray}\label{EqN25}
&&\hspace*{-3.5mm}  p_{D\pm} (T, \nu, m) \equiv \pm T g \hspace*{-1.5mm} \int \hspace*{-1.5mm}  \frac{d^D k }{(2 \pi \hbar)^D}   \ln  \left[1 \pm \exp\left[ \frac{\nu - e_m(k)}{T} \right] \right] , ~ 
\rm where\quad e_m (k) \equiv  \sqrt{m^2+ k^2}   , 
\end{eqnarray}
one can generalize the BE-FD  decomposition identity (\ref{EqB11}) to the  dimension $D$ 
for the fractional mass and charge values 
\begin{eqnarray}\label{EqN26}
p_{D-}\left( T, \nu, m  \right) \equiv & 2^D p_{D-}\left( T, \frac{\nu}{2}, \frac{m}{2} \right) -  
2^D p_{D+}\left( T, \frac{\nu}{2}, \frac{m}{2} \right) . 
\end{eqnarray} 

For chargeless and massless particles, i.e. for $\nu=0$ and $m=0$,  the BE-FD decomposition identity   (\ref{EqN26})  gives us the following relation
between the BE and FD momentum integrals 
\begin{eqnarray}\label{EqN27}
&& \hspace*{-3.5mm}
p_{D-}\left( T, 0, 0 \right) \equiv \frac{2^D}{2^D-1} p_{D+}\left( T, 0, 0 \right) 
  \quad \Rightarrow \quad   \int\limits_0^\infty \frac{x^D \, dx}{e^x -1} = \frac{2^D}{2^D-1} \int\limits_0^\infty \frac{x^D \, dx}{e^x +1}  \, , 
\end{eqnarray}
which for $D=3$ leads to a well-known identity
\begin{eqnarray}\label{EqN28}
&& \hspace*{-3.5mm}
   \int\limits_0^\infty \frac{x^3 \, dx}{e^x -1}  = \frac{8}{7} \int\limits_0^\infty \frac{x^3 \, dx}{e^x +1} = \frac{\pi^4}{15}  \, , 
\end{eqnarray}
Note, however, that 
 the right equation  (\ref{EqN27})  follows from the left one  after  integrating the pressures of massless and chargeless particles  over the angles, first,  and, then, after  integrating  them over $d k^D$ by parts. 

Applying the identity  (\ref{EqN26})  to its right hand side $n$ times, one obtains another identity
\begin{eqnarray}\label{EqN29}
p_{D-}\left( T, \nu, m  \right) \equiv & 2^{D n}  p_{D-}\left( T, \frac{\nu}{2^n}, \frac{m}{2^n} \right) -  
\sum\limits_{k=1}^n 2^{D k} p_{D+}\left( T, \frac{\nu}{2^k}, \frac{m}{2^k} \right) . 
\end{eqnarray} 
For $n \gg \ln \left[ \max (\frac{\nu}{T}; \frac{m}{T}) \right]$ with the help of identity  (\ref{EqN27})  one can establish an approximative relation 
\begin{eqnarray}\label{EqN30}
p_{D-}\left( T, \nu, m  \right) \simeq &   \frac{2^{D(n+1)}}{2^D-1}  p_{D+}\left( T, \frac{\nu}{2^n}, \frac{m}{2^n} \right) -  
\sum\limits_{k=1}^n 2^{D k} p_{D+}\left( T, \frac{\nu}{2^k}, \frac{m}{2^k} \right) ,
\end{eqnarray} 
which again relates the pressures of  BE  and FD particles.  Note that  Eqs. (\ref{EqN25}), (\ref{EqN26}), (\ref{EqN29}) and (\ref{EqN30})   are valid for the particles with the hard-core repulsion, i.e. for $\nu = \mu - 2^{(D-1)} V_D \, p_{D-}\left( T, \nu, m  \right)$, where the eigenvolume of particles in the $D$-dimensional space is  denoted as $V_D$.

\section{Conclusions}

In this work we recapitulate the VdW  equation of state of BE particles with the hard-core repulsion in the grand canonical ensemble. 
Our analysis  shows that the pressure of  non-relativistic BE particles is mathematically equivalent  to the one of the exactly solvable model with the 1-st order PT known as the sSMM. The EoS of BE particles  corresponds to the sSMM with  the  vanishing surface tension coefficient and the Fisher exponent $\tau_F = \frac{5}{2}$. 
Such a similarity  allows us to show that within the present approach  the high density phase of BE particles is a classical macro-cluster  with vanishing entropy at any temperature which, similarly to the classical hard spheres,  is a kind of solid state. Considering  the limit of very small eigenvolume  of BE particles we argue that the ideal gas of  BE particles has the  1-st order  PT as it was suggested by K. Huang in his famous textbook \cite{Huang} a long time ago.

To explicitly demonstrate that a macro-cluster with the BE statistics does not exist in this EoS  we investigate some peculiar relations between the pressures of BE and FD particles, the BE-FD decomposition identities,  showing that under some conditions the pressure of FD particles can be identically  rewritten in terms of two  BE pressures. Moreover, we establish an exact representation of  the pressure of BE particles of mass $m$, charge $1$
and degeneracy $g$ as a series of pressures
of FD particles with  the masses $2^k m$, the charge $2^k$ and degeneracy  $2^{-3 k}g$, where $k$ are positive natural numbers. These new  relations  help us to correctly  interpret the properties of  a high density phase of BE particles with hard-core repulsion.  

 In fact, here we establish a principally new look at the problem of BE condensation. Of course, the considered model is oversimplified, but now one can use all the achievements of the SMM \cite{Bondorf,SMM0,KABsmm1,KABsmm2,KABsmm3}
and introduce the surface part  $\sigma(T) k^\frac{2}{3}$ of  the  free energy of  $k$-particle clusters (here $\sigma(T)$ is the temperature dependent coefficient of surface tension). 
Such a modification will make model more realistic, since  the surface part of free energy partly accounts for  the 
short range attraction among the constituents like it is done in the full SMM. Note that in 
this case, however,  the modified right hand side of Eq.  (5) cannot be already reduced to the pressure of
point-like particles  $p_- (T, \nu)$ with the shifted chemical potential  
$\nu  = \mu - 4 V_0 p_- $.

\vspace*{2.2mm}

{\bf Acknowledgements.} The authors thank Oleksandr Vitiuk, Oleksandr  Khasai, Maksim Tomchenko,  Nazar Yakovenko and Gennady Zinovjev for 
fruitful discussions. 
The authors are thankful to Edward Gorbar for reading the manuscript and for his valuable  critique. 
The work of K.A.B., O.I.I. and B.E.G. was supported in part by the Program of Fundamental Research in High Energy and Nuclear Physics launched by the Section of Nuclear Physics of the National Academy of Sciences of Ukraine. K.A.B. is  grateful to the COST Action CA15213 ``THOR`` for supporting  his   networking.

\end{document}